\documentclass[conference]{IEEEtran}
\usepackage{amsmath,amsfonts,amssymb}
\usepackage{array}
\usepackage{bm}
\usepackage{textcomp}
\usepackage{stfloats}
\usepackage{url}
\usepackage{verbatim}
\usepackage{graphicx}
\usepackage{cite}
\usepackage{color}
\usepackage{optidef}
\usepackage{physics}
\usepackage[caption=false, font=footnotesize]{subfig}
\usepackage{algorithm}
\usepackage{algpseudocode}
\usepackage{etoolbox}
\usepackage{mathtools}
\usepackage{setspace}
\usepackage{multirow}
\usepackage{booktabs}
\usepackage{threeparttable}
\usepackage{makecell}

\graphicspath{{./plots/prod0103/}}
\newcommand{\Tref}[1]{Table~\ref{#1}}
\newcommand{\Fref}[1]{Fig.~\ref{#1}}
\newcommand{\Sref}[1]{Section~\ref{#1}}

\newcommand{\Nsk}{\(N_{\mathrm{sk}}\)}

\DeclareMathOperator*{\argmax}{argmax}
\newcommand\blfootnote[1]{%
  \begingroup
  \renewcommand\thefootnote{}\footnote{#1}%
  \addtocounter{footnote}{-1}%
  \endgroup
}

\begin{document}

\title{\huge Exploiting Skyrmions in Free-Space Optical Communication}

\author{
\IEEEauthorblockN{Ryosuke Hara, Satoshi Iwamoto, and Shinya~Sugiura$^*$}
\IEEEauthorblockA{Institute of Industrial Science, The University of Tokyo\\
E-mail: \{hararyo, iwamoto, sugiura\}@iis.u-tokyo.ac.jp 
}

    \thanks{The authors are with the Institute of Industrial Science, The University of Tokyo, Tokyo 153-8505, Japan (e-mail: \{hararyo, iwamoto, sugiura\}@iis.u-tokyo.ac.jp). (\textit{Corresponding author: Shinya Sugiura}.)}
    \thanks{This work was supported in part by the Japan Science and Technology Agency (JST) FOREST (Grant JPMJFR2127), in part by the JST ASPIRE (Grant JPMJAP2345), in part by the JST CREST (JPMJCR19T1), and in part by the Japan Society for the Promotion of Science (JSPS) KAKENHI (Grant 24K21615). Part of this paper was submitted for presentation at IEEE ICC 2026, Glasgow, U.K..}
\vspace*{-6mm}
}

\markboth{}
{Shell \MakeLowercase{\textit{et al.}}: A Sample Article Using IEEEtran.cls for IEEE Journals}

\IEEEpubid{}

\maketitle

\begin{abstract}
    In this paper, we propose a novel free-space optical (FSO) communication system utilizing optical skyrmions. We introduce a scheme referred to as skyrmion number modulation (SkM), which employs index modulation by encoding information onto the skyrmion number, a topological invariant preserved during free-space propagation. This topological nature offers the potential for inherent robustness against atmospheric turbulence-induced wavefront distortions, which limit the performance of conventional FSO systems. More specifically, we demonstrate that the fluctuation of the received skyrmion number is mitigated by a proposed intensity-based masking technique. Finally, our performance analysis based on a discrete memoryless channel framework confirms that the proposed system exhibits near-ideal robustness under weak turbulence and supports high-order modulation in moderate regimes.
\end{abstract}

\section{Introduction}%
\label{sec:Introduction}
\blfootnote{Preprint for publication in \textit{IEEE International Conference on Communications Workshops (ICC Workshops)}, Glasgow, Scotland, UK, May 2026. 
$\copyright$ 2026 IEEE. Personal use of this material is permitted. Permission from IEEE must be obtained for all other uses, in any current or future media, including reprinting/republishing this material for advertising or promotional purposes, creating new collective works, for resale or redistribution to servers or lists, or reuse of any copyrighted component of this work in other works.}
\IEEEPARstart{S}{kyrmions} are topologically stable quasiparticles characterized by swirling vector field configurations~\cite{Shen2024-cg}.
Along with recent progress in structured light~\cite{Forbes2021-aw}, optical skyrmions have been realized in guided waves and paraxial beams, leveraging various degrees of freedom. 
Optical skyrmionic beams and paraxial optical skyrmions exhibit spatially varying polarization patterns that cover all polarization states, and the corresponding normalized Stokes vectors form a skyrmionic structure~\cite{Gao2020-as}.
The skyrmion number $N_{\mathrm{sk}}$ is a topological invariant integer that characterizes skyrmions and may represent a robust information carrier in optical communications~\cite{Shen2024-cg,Wang2024-af,Wang2025ssfmturb}, owing to its conservation during free-space propagation~\cite{Gao2020-as} and anticipated topological robustness.
More specifically, $N_{\mathrm{sk}}$ represents the number of times the vector field wraps around a unit sphere, and is invariant under continuous deformations~\cite{Nagaosa2013-bz,Piette1995-jq}, suggesting inherent resilience against perturbations.
Although research on optical skyrmions is still in its infancy, recent studies have demonstrated this resilience experimentally and numerically, showing that $N_{\mathrm{sk}}$ is preserved even when the local skyrmion density structure is scrambled~\cite{Wang2024-af}.
Under atmospheric turbulence, experiments and numerical simulations utilizing the split-step Fourier method (SSFM)~\cite{Schmidt2010-bk,Peters2025-io} have further confirmed the robustness of $N_{\mathrm{sk}}$, though deviations occur under strong turbulence conditions~\cite{Wang2025ssfmturb}.
Furthermore, optical skyrmions are described solely by linear Maxwell's equations~\cite{Gutierrez-Cuevas2021-go}, and this linearity facilitates the exploitation of an unbounded $N_{\mathrm{sk}}$ range, in contrast to the limited integer range of magnetic skyrmions.

In wireless communications, free-space optical (FSO) communication employing orbital angular momentum (OAM)~\cite{Allen1992-ru} has been extensively studied as a means of enhancing data rate, though its achievable performance is limited by atmospheric turbulence.
More specifically, random fluctuations of the refractive index distort the wavefront~\cite{Andrews2005-ro}, thus degrading channel quality due to scintillation and mode crosstalk\cite{Amhoud2020-ye}.
These detrimental effects were analyzed and evaluated in the experiments~\cite{van-Iersel2023-fh} and the numerical simulations using SSFM~\cite{Amhoud2020-ye}.
The inherent stability of skyrmions suggests the possibility of robust data transmission without relying on complicated techniques.
However, an FSO communication system utilizing optical skyrmions has not yet been proposed.
Furthermore, the detailed statistical behavior and communication performance of $N_{\mathrm{sk}}$ in turbulent channels remain to be investigated.

To efficiently convey information via the band-limited channel, a sophisticated modulation technique plays an important role~\cite{hanzo2004quadrature}. A recent index modulation (IM) concept is designed to exploit available communication resources by activating a subset of legitimate symbol indices, in addition to conventional modulation schemes such as phase-shift keying and quadrature-amplitude modulation~\cite{Sugiura2017-xe}. To be more specific, there are several IM architectures, supporting the dimensions of the space~\cite{di2013spatial}, time~\cite{nakao2016single}, frequency~\cite{ishikawa2016subcarrier}, and beam direction~\cite{sugiura2014coherent}. Note that IM schemes operating in the frequency and time domains have been shown to achieve superior bit-error rate (BER) performance compared to traditional methods, particularly in low-rate scenarios\cite{Sugiura2017-xe}.

The novel contribution of this paper is to propose an FSO communication system that utilizes the skyrmion number for IM, referred to as \textit{skyrmion number modulation (SkM)}. We introduce a novel intensity-based masking technique at the receiver to improve the robustness of $N_{\mathrm{sk}}$ calculations. Moreover, we derive the theoretical channel capacity, symbol-error rate (SER), and BER of the proposed scheme, and we optimize the SkM constellation to maximize the derived channel capacity. We evaluate SkM in a realistic FSO channel based on SSFM in a comprehensive manner. More specifically, we characterize the statistical behavior of the skyrmion number under atmospheric turbulence. Finally, we demonstrate the achievable capacity, SER, and BER of the proposed scheme.

\section{Preliminaries}%
\label{sec:Preliminaries}
\subsection{Optical Skyrmionic Beams}%
\label{subsec:skyrmionic_beams}
Since the optical skyrmionic beam is categorized into the specific class of vector beams, the electric field is modeled as a superposition of two orthogonal spatial modes, which are coupled with orthogonal polarization states as follows \cite{Nape2022-dk,Gao2020-as}:
\begin{IEEEeqnarray}{rCL}%
    \label{eq:vector_beam}
    \ket{\Psi(\bm{r})} &=& u_0 (\bm{r}) \ket{0} + e^{i\theta_0} u_1 (\bm{r}) \ket{1},
\end{IEEEeqnarray}
where $\ket{0}$ and $\ket{1}$ denote the orthogonal polarization basis (e.g., right- and left-circular polarizations $\ket{R}$ and $\ket{L}$), $\bm{r} \in \mathbb{R}^3$ is the spatial coordinate, $u_{0} (\bm{r})$ and $u_{1} (\bm{r})$ are the orthogonal spatial modes, $i$ is the imaginary unit, and $\theta_{0}$ is a constant global phase difference.

The polarization state at any point is represented by the Stokes vector $\mathbf{s}(\bm{r})$\cite{Gao2020-as,Martinelli2017-jh}:
    $\mathbf{s}(\bm{r}) = \bra{\Psi(\bm{r})} \boldsymbol{\sigma} \ket{\Psi(\bm{r})}$,
where $\boldsymbol{\sigma} = (\sigma_{x}, \sigma_{y}, \sigma_{z})$ is the vector of Pauli matrices defined as
    $\sigma_{x} = \ket{0}\bra{1} + \ket{1}\bra{0}$,
    $\sigma_{y} = i \ket{1}\bra{0} - i \ket{0}\bra{1}$, and
    $\sigma_{z} = \ket{0}\bra{0} - \ket{1}\bra{1}$\cite{Ye2024-hq}.
As the orientation of the vector is of primary interest, we employ the normalized Stokes vector, which is given by
    $\mathbf{S}(\bm{r}) = {\mathbf{s}(\bm{r})}/{\| \mathbf{s}(\bm{r})\|}$,
where $\|\cdot\|$ denotes the Euclidean norm\footnote{$\mathbf{S}(\bm{r})$ is defined on the rotated Poincaré sphere $\mathbb{S}^2$\cite{Ye2024-hq}.
    When the basis is chosen as $\ket{0} = \ket{R}$ and $\ket{1} = \ket{L}$, $\mathbf{S}$ corresponds directly to the conventional Stokes parameters $(S_1, S_2, S_3)$\cite{Martinelli2017-jh}.}.
This normalization is equivalent to division by the local intensity $I (\bm{r}) = S_0 = \braket{\Psi(\bm{r})}$\cite{Martinelli2017-jh} under fully polarized conditions.

\begin{figure}[!t]
    \centering
    \subfloat[]{\includegraphics[width=0.45\linewidth]{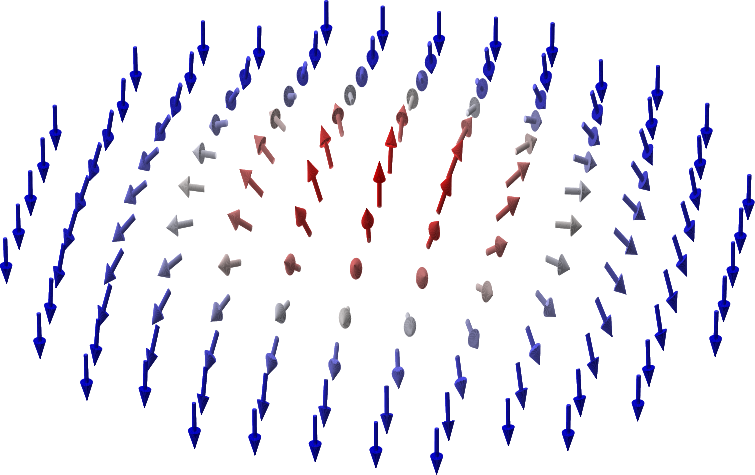}\label{fig:skyrmions_a}}
    \hfill
    \subfloat[]{\includegraphics[width=0.45\linewidth]{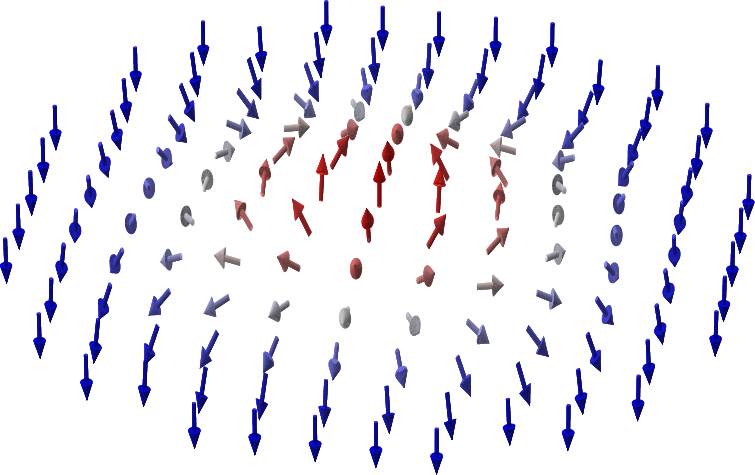}\label{fig:skyrmions_b}}
    \\
    \subfloat[]{\includegraphics[width=0.45\linewidth]{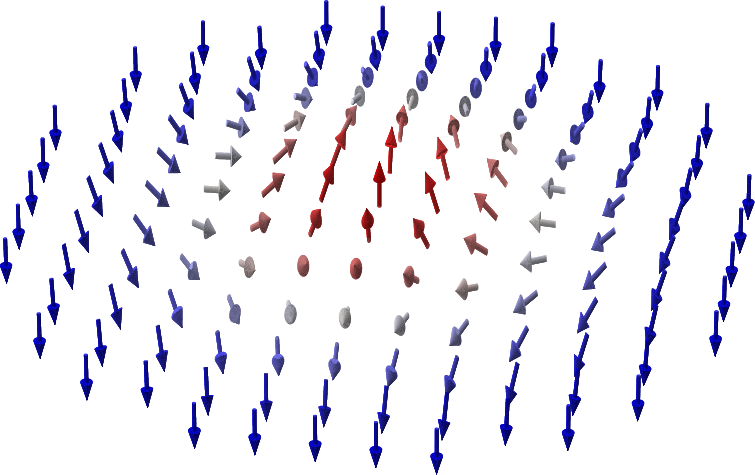}\label{fig:skyrmions_c}}
    \hfill
    \subfloat[]{\includegraphics[width=0.45\linewidth]{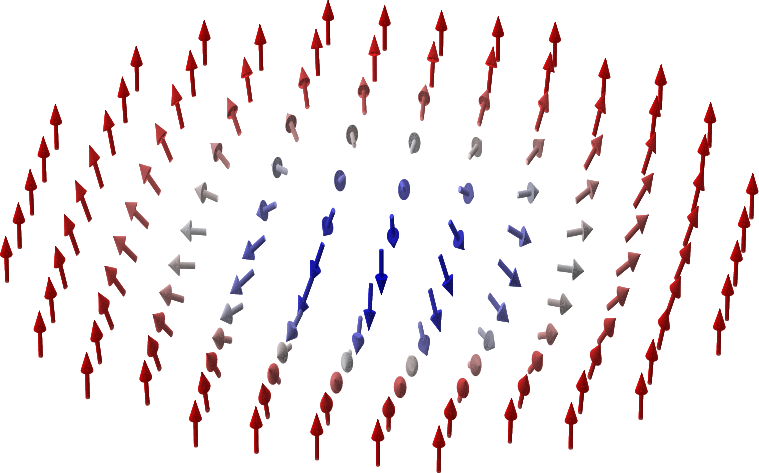}\label{fig:skyrmions_d}}
    \caption{Vector field configurations of skyrmion beams.
    (a) N\'{e}el-type skyrmion with $N_{\mathrm{sk}} = 1$.
    (b) Higher-order skyrmion with $N_{\mathrm{sk}} = 2$.
    (c) Anti-skyrmion with $N_{\mathrm{sk}} = -1$.
    (d) N\'{e}el-type skyrmion with opposite polarity, yielding $N_{\mathrm{sk}} = -1$.}%
    \label{fig:skyrmions}
\end{figure}

\Fref{fig:skyrmions} illustrates the topologically non-trivial polarization textures of optical skyrmions.
These textures typically feature a continuous transition of the polarization state from one pole of the Poincaré sphere at the beam center to the antipodal pole at the periphery, while the polarization orientation rotates as one traverses the azimuthal coordinate around the beam center\cite{Gao2020-as}.
The topological charge of this texture is quantified by the skyrmion number $N_{\mathrm{sk}}$, which counts the number of times the vector field wraps around the Poincaré sphere.
Assuming beam propagation along the $+z$ axis and considering the transverse plane coordinates $\bm{\rho} = (x, y)$, $N_{\mathrm{sk}}$ is calculated by an integral over the entire plane $\mathcal{R}$ as follows~\cite{Nagaosa2013-bz}:
\begin{IEEEeqnarray}{rCL}%
    \label{eq:Skyrmion_number}
    N_{\mathrm{sk}} & = & \frac{1}{4\pi}
    \int_{\mathcal{R}} \mathbf{S} (\bm{\rho}) \cdot
    \left(\frac{\partial\mathbf{S} (\bm{\rho})}{\partial{x}} \times
    \frac{\partial\mathbf{S} (\bm{\rho})}{\partial{y}} \right)
    dxdy.
\end{IEEEeqnarray}

Typically, skyrmionic beams are generated by using Laguerre-Gaussian (LG) modes as the orthogonal spatial modes. The complex-valued amplitude of an LG mode in the cylindrical coordinates $(r, \varphi, z)$ is given by~\cite{Yao2011-sd}
\begin{IEEEeqnarray}{rCL}%
    \label{eq:LG_mode}
    u_p^\ell(r, \varphi, z)
    & = & \sqrt{\frac{2p!}{\pi(p+|\ell|)!}} \frac{1}{w (z)}
    \left[\frac{r\sqrt{2}}{w (z)}\right]^{|\ell|}
    \exp\left(-\frac{r^2}{w^2 (z)} \right) \nonumber\\
    && \hspace{-1.0em} \times L_p^{|\ell|}\left(\frac{2r^2}{w^2(z)} \right)
    \exp(i\ell\varphi)
    \exp\left(\frac{ikr^2z}{2(z^2 + z_R^2)} \right) \nonumber\\
    && \hspace{-1.0em} \times \exp\left(-i(2p+|\ell|+1) \tan^{-1}\left(\frac{z}{z_R}\right) \right),
\end{IEEEeqnarray}
where $p$ and $\ell$ are the radial and azimuthal indices, respectively, $L_{p}^{|\ell|}(\cdot)$ is the generalized Laguerre polynomial, $w(z) = w_0 \sqrt{1 + (z/z_R)^2}$ is the beam width, and $z_{R} = \pi w_{0}^{2}/\lambda$ is the Rayleigh range with the beam waist $w_{0}$ at the focus $z =0$. Also, $\lambda$ is the wavelength and $k = 2\pi/\lambda$ is the wavenumber.

Substituting two LG modes $u_0 = u_0^{\ell_0}$ and $u_1 = u_0^{\ell_1}$ for $|\ell_0| \neq |\ell_1|$ into \eqref{eq:vector_beam} yields a skyrmionic beam, whose \Nsk~remains invariant during propagation\cite{Gao2020-as}. In this configuration, \Nsk~is uniquely determined by the azimuthal indices. By simplifying the analytical expression derived in\cite{Gao2020-as}, we obtain the following concise relationship of
\begin{IEEEeqnarray}{rCL}%
    \label{eq:Nsk_simplified}
    N_{\mathrm{sk}} = \mathrm{sgn}(|\ell_{1}| - |\ell_{0}|)(\ell_{1} - \ell_{0}).
\end{IEEEeqnarray}
Here, the sign term $\mathrm{sgn}(|\ell_{1}| - |\ell_{0}|)$ dictates the skyrmion's polarity, i.e., orientation of the central polarization, while the difference $\ell_{1} - \ell_{0}$ determines the vorticity, i.e.,  azimuthal winding number. Specifically, if $\ell_{1} > \ell_{0}$, the polarization rotation follows the azimuthal angle, resulting in positive vorticity (Fig.~\ref{fig:skyrmions}(a)); by contrast, the relationship of $\ell_{1} < \ell_{0}$ yields negative vorticity, corresponding to an anti-skyrmion (Fig.~\ref{fig:skyrmions}(c))\cite{Shen2024-cg}.

\subsection{Atmospheric Turbulence}%
\label{subsec:atmospheric_turbulence}
Atmospheric turbulence manifests as random fluctuations in the refractive index, arising from spatial inhomogeneities in temperature, humidity, and pressure.
Since an optical beam propagates through this turbulent medium, it accumulates random phase perturbations that distort the wavefront. To mathematically model these refractive index fluctuations, a spatial power spectrum is commonly employed.
In this paper, we employ the modified Kolmogorov spectrum \cite{Andrews2005-ro}.
\begin{IEEEeqnarray}{rCL}
    \Phi_n(\kappa) & = & 0.033 C_n^2 \left[ 1 + 1.802\left(\frac{\kappa}{\kappa_l}\right) - 0.254{\left(\frac{\kappa}{\kappa_l}\right)}^{7/6} \right] \nonumber \\
    && \times \frac{\exp(-\kappa^2/\kappa_l^2)}{{(\kappa^2 + \kappa_0^2)}^{11/6}},%
\end{IEEEeqnarray}
where $\kappa$ is the spatial frequency, $\kappa_l = 3.3/l_0$, $\kappa_0 = 2\pi/L_0$, $C_{n}^{2}$ is the refractive index structure parameter, $l_{0}$ is the inner scale, and $L_0$ is the outer scale of turbulence.

The turbulence condition is characterized by several key parameters~\cite{Peters2025-io}.
For example, $C_{n}^{2}$ is the most fundamental parameter, typically ranging from $10^{-17} \, \mathrm{m}^{-2/3}$ to $10^{-13} \, \mathrm{m}^{-2/3}$\cite{Karp2013-hw}, where higher values indicate stronger turbulence intensities. However, $C_{n}^{2}$ represents a local strength and does not fully characterize the cumulative distortion over the entire propagation path.
Hence, the turbulence strength over a specific propagation length $L$ is typically represented by the Rytov variance as follows:
    $\sigma_{R}^{2} = 1.23 \, C_{n}^{2} \, k^{7/6} \, L^{11/6}$ \cite{Peters2025-io}.
The turbulence regime is classified to be weak for $\sigma_{R}^{2} < 1$, moderate for $\sigma_{R}^{2} \approx 1$ and strong for $\sigma_{R}^{2} > 1$\cite{Andrews2005-ro}.\footnote{In the context of vector beams, atmospheric turbulence primarily manifests as distortions in the constituent spatial modes, potentially deforming the polarization profile\cite{Nape2022-dk}.
Physically, the effect of turbulence $\mathcal{T}$ on the polarization basis states is negligible and can be dismissed.
However, it induces crosstalk among the spatial modes, which may be expressed as
    $\mathcal{T}(u_p^{\ell} \ket{0}) = \sum_{p', \ell'} c_{p', \ell'} u_{p'}^{\ell'} \ket{0}$,
where $c_{p', \ell'} \in \mathbb{C}$ are complex-valued coefficients satisfying the normalization condition $\sum_{p', \ell'} |c_{p', \ell'}|^2 = 1$.
Consequently, the local amplitude ratio and relative phase between the orthogonal components ($\ket{0}$ and $\ket{1}$ terms in~\eqref{eq:vector_beam}) are perturbed spatially, thereby distorting the polarization texture.}

\section{System Model}%
\label{sec:System_Model}

\subsection{Skyrmion Number Modulation}%
\label{subsec:SkM_Transmitter}
In the proposed SkM scheme, information bits are modulated onto the skyrmion number $N_{\mathrm{sk}}$. This is motivated by the IM principle, since information is conveyed by activating a single skyrmion number from a predefined candidate set.
Let $\mathcal{C} = \{n_1, n_2, \dots, n_M\}$ be the constellation of $M$ skyrmion numbers, where $M = 2^K$ is the constellation size. Hence, each activation of skyrmion number conveys $K$ information bits. Here, the skyrmion numbers in the constellation are distinct integers, which are sorted in ascending order, i.e., $n_1 < n_2 < \cdots < n_M$.

Then, based on the activated skyrmion number, the transmitter generates a corresponding skyrmionic beam $\ket{\Psi_{N_\textrm{sk}}}$, carrying the skyrmion number $N_{\mathrm{sk}}$. In this paper, we employ the superposition of orthogonal LG modes described in \Sref{subsec:skyrmionic_beams}. For simplicity, we consider the configuration, where the azimuthal indices are set to $\ell_0 = 0$ and $\ell_1 = N_\textrm{sk}$ with orthogonal circular polarization states $\ket{R}$ and $\ket{L}$. Accordingly, the transmitted field is formulated by
\begin{IEEEeqnarray}{rCL}
    \label{eq:ni_skyrmion_beam}
    \ket{\Psi_{N_\textrm{sk}}(\bm{r})} = \frac{1}{\sqrt{2}}u_0^0(\bm{r}) \ket{R} + \frac{1}{\sqrt{2}}u_0^{N_\textrm{sk}}(\bm{r}) \ket{L}.
\end{IEEEeqnarray}

\subsection{Mask-Aided Skyrmion Number Calculation}%
\label{subsec:Receiver_Masking}
The transmitted beam propagates through an atmospheric turbulence channel with a distance $L$ between the transmitter and the receiver.
Representing the atmospheric turbulence as an operator $\mathcal{T}$, the light field at the receiver plane ($z = L$) is expressed as
\begin{IEEEeqnarray}{rCL}
    \label{eq:turbulent_light_field}
    \ket{\Psi_\text{turb}(\bm{\rho})} &=& \mathcal{T}(\ket{\Psi_{N_\textrm{sk}}(\bm{\rho})}) \nonumber\\
    &=& \sum_{p, \ell} u_{p}^{\ell}(\bm{\rho}) \left( c_{p, \ell} \ket{R} + d_{p, \ell} \ket{L} \right),
\end{IEEEeqnarray}
where $c_{p, \ell}, d_{p, \ell} \in \mathbb{C}$ are complex-valued coefficients, satisfying $\sum_{p, \ell} (|c_{p, \ell}|^2 + |d_{p, \ell}|^2) = 1$.

The detection process begins by measuring the spatial distribution of the Stokes vector across the aperture. This typically requires intensity measurements, projected onto different polarization bases.
Accounting for thermal and shot noises, which are inherent in photodetection, the measured Stokes vector $\tilde{\mathbf{s}}(\bm{\rho})$ and its normalized form $\tilde{\mathbf{S}}(\bm{\rho})$ are represented, respectively, by
\begin{IEEEeqnarray}{rCL}
    \label{eq:stokes_measurement}
    \tilde{\mathbf{s}}(\bm{\rho}) &=& \bra{\Psi_\text{turb}(\bm{\rho})} \boldsymbol{\sigma} \ket{\Psi_\text{turb}(\bm{\rho})} + \bm{\eta}(\bm{\rho}) \\
    \tilde{\mathbf{S}}(\bm{\rho}) &=& \frac{\tilde{\mathbf{s}}(\bm{\rho})}{\|\tilde{\mathbf{s}}(\bm{\rho})\|},
    \label{eq:stokes_normalized_measured}
\end{IEEEeqnarray}
where $\bm{\eta}(\bm{\rho}) \in \mathbb{R}^3$ denotes the additive measurement noise.

The receiver calculates the activated skyrmion number according to the definition~\eqref{eq:Skyrmion_number}.
However, this direct calculation faces a limitation because it relies on the normalized vector $\tilde{\mathbf{S}}(\bm{\rho})$, which inherently disregards the intensity profile of the received beam. In the low-intensity region, the signal tends to be contaminated by the additive noise $\bm{\eta}(\bm{\rho})$ and the turbulence-induced crosstalk, described in~\eqref{eq:turbulent_light_field}. Furthermore, the normalization step in~\eqref{eq:stokes_normalized_measured} scales disturbance-dominated inputs into unit-length vectors with random directions. Such erratic vectors may corrupt the integration, leading to a significant deviation between the calculated $N_{\mathrm{sk}}$ and the transmitted skyrmion number~\cite{Wang2025ssfmturb}.

To mitigate the erosive effects, we introduce an \textit{intensity-based masking technique}.
More specifically, we apply a weighting function $W(\bm{\rho}) \in [0,1]$ to suppress the effects of the unreliable low-intensity regions. The masked skyrmion number $\tilde{N}$ is computed as
\begin{IEEEeqnarray}{rCL}
    \label{eq:masked_Nsk}
    \tilde{N} & = & \frac{1}{4\pi}
    \int_{\mathcal{A}} W(\bm{\rho}) \tilde{\mathbf{S}} (\bm{\rho}) \cdot
    \left(\frac{\partial\tilde{\mathbf{S}} (\bm{\rho})}{\partial{x}} \times
    \frac{\partial\tilde{\mathbf{S}} (\bm{\rho})}{\partial{y}} \right)
    dxdy,
\end{IEEEeqnarray}
where $\mathcal{A}$ is the integration area of the receiver aperture. While $N_{\mathrm{sk}}$ is theoretically an integer, the calculated $\tilde{N}$ is a continuous random variable due to the stochastic nature of the channel and the residual noise.
While in the experimental studies~\cite{Wang2024-af}, the low-intensity pixels are heuristically excluded from the received signals, we herein formulate our masking technique as a standard receiver component, which is referred to as the \textit{scaled-mean binary mask}, whose detail is introduced and discussed in \Sref{sec:Channel_characterization}.

\subsection{Maximum-Likelihood Symbol Detection}
Finally, the receiver estimates the transmitted integer skyrmion number $\hat{N}$ from the calculated continuous value $\tilde{N}$ by hard decision. Hence, the hard-decision boundaries $\mathcal{B} = \{\beta_1, \beta_2, \dots, \beta_{M-1}\}$ have to be designed, where $\beta_k$ represents the threshold separating adjacent skyrmion numbers $n_k$ and $n_{k+1}$ $(1\le k\le M-1)$ in the constellation considered.
Based on these boundaries, the continuous skyrmion number $\tilde{N}$ calculated at the receiver is partitioned into $M$ disjoint decision regions
    $\mathcal{D}_k = \{ x \in \mathbb{R} \mid \beta_{k-1} < x \le \beta_k \}$,
where we assume $\beta_0 = -\infty$ and $\beta_M = +\infty$. The skyrmion number is hard-decided as $\hat{N} = n_k$ if $\tilde{N} \in \mathcal{D}_k$.

In this paper, the decision boundaries $\mathcal{B}$ are optimized to minimize the pairwise error probability between adjacent symbols. Under the assumption that all skyrmion numbers are activated with equal probability, our decision criterion corresponds to maximum-likelihood detection.
Therefore, we obtain the optimal boundaries $\beta_k$ given by the intersection of two conditional probability density functions (PDFs) associated with adjacent skyrmion numbers as follows:
    $f(\beta_k | n_k) = f(\beta_k | n_{k+1})$,
where $f(x | n_k)$ denotes the PDF of the received value $\tilde{N}$ given that the skyrmion number $n_k$ is activated.

\section{Channel Characterization}%
\label{sec:Channel_characterization}

\subsection{Simulation Framework}%
\label{subsec:sim_setup}
\subsubsection{System Specifications and Turbulence Parameters}
The system parameters adopted in this paper are listed in \Tref{tab:simulation_params}. Similar to the conventional study of the OAM-based FSO system~\cite{Amhoud2020-ye}, our system operates at a wavelength of $\lambda=850~\mathrm{nm}$ over a propagation distance of $L=1~\mathrm{km}$ with a transmitter beam waist of $w_0=1.6~\mathrm{cm}$.
We consider the set of legitimate skyrmion numbers as $\mathcal{S} = \{-8, \dots, -1, 1, \dots, 8\}$,
where a specific constellation with the size $M$ is constructed as subsets of $\mathcal{S}$. According to~\cite{Amhoud2020-ye}, we assume a sufficiently large receiver aperture diameter $d_{\mathrm{rx}}$. More specifically, given that the spot size of LG modes comprising skyrmionic beams scales as $w(z)\sqrt{2p+|\ell|+1}$~\cite{Phillips1983-fk}, we set $d_{\mathrm{rx}} = 2w(L)\sqrt{\ell_{\max}+1} \approx 13.97~\mathrm{cm}$ based on the maximum azimuthal index $\ell_{\max}=8$. 
Furthermore, we consider three refractive-index constants, i.e., $C_n^2 = 1.0\times10^{-15}$, $2.5\times10^{-14}$, and $1.0\times10^{-13}~\mathrm{m}^{-2/3}$. These values correspond to Rytov variances of $\sigma_R^2 = 0.04, 1.0$, and $4.0$, exhibiting the weak, moderate, and strong turbulence regimes, respectively.

\begin{table}[tb]
    \caption{Simulation Parameters}%
    \label{tab:simulation_params}
    \centering
    \begin{threeparttable}
        \scriptsize
        \begin{tabular}{@{}l c l@{}}
            \toprule
            Parameter                                    & Notation             & Value                               \\
            \midrule
            Wavelength                                   & $\lambda$            & 850 nm                              \\
            Propagation distance                         & $L$                  & 1 km                                \\
            Beam waist at transmitter                    & $w_0$                & 1.6 cm                              \\

            Range of skyrmion numbers $N_{\mathrm{sk}}$  &                      & $-8, \dots, -1,1, \dots, 8$         \\
            Receiver aperture diameter                   & $d_{\mathrm{rx}}$    & 13.97 cm                            \\
            \midrule
            Refractive index structure constant & $C_n^2$              & \makecell[l]{$1.0 \times 10^{-15}$, \\ $2.5\times 10^{-14}$, \\ $1.0 \times 10^{-13}~\mathrm{m}^{-2/3}$} \\

            Rytov variance                      & $\sigma_R^2$         & 0.04, 1.0, 4.0                      \\

            Inner scale                                  & $l_0$                & 5 mm                                \\
            Outer scale                                  & $L_0$                & 20 m                                \\
            \midrule
            Number of realizations                       &                      & 20,000                              \\
            Grid size                                    & $M_g$                & $1024 \times 1024$                  \\
            Grid spacing                                 & $\Delta x, \Delta y$ & 0.733 mm                            \\
            Number of phase screens                      &                      & 20                                  \\
            Phase screen spacing                         & $\Delta z$           & 50 m                                \\
            \bottomrule
        \end{tabular}

    \end{threeparttable}
\end{table}

\subsubsection{Split-Step Fourier Method}
\begin{figure}
    \centering
    \includegraphics[width=\linewidth]{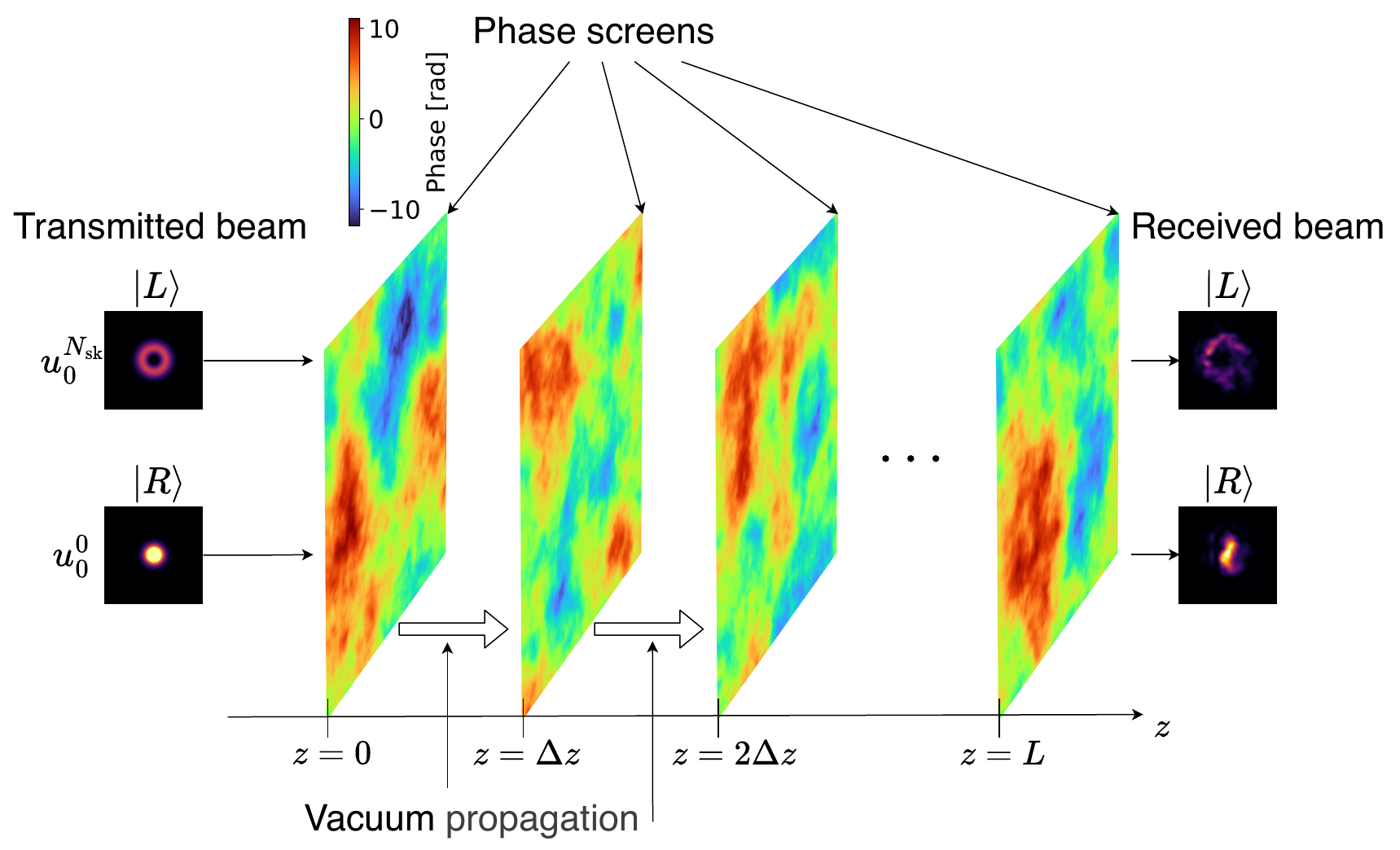}
    \caption{Schematic of beam propagation simulation using SSFM.}%
    \label{fig:split_step}
\end{figure}

We employ SSFM to numerically compute the turbulent light field $\ket{\Psi_{\text{turb}}(\bm{\rho})}$ at the receiver plane.
As illustrated in \Fref{fig:split_step}, the propagation path is divided into multiple discrete steps.
In each segment, the beam undergoes vacuum diffraction, computed efficiently via the fast Fourier transform (FFT) based on Fourier optics~\cite{Schmidt2010-bk}, followed by transmission through a random phase screen representing the accumulated turbulence. For the phase screen generation, a simple FFT-based method is employed, similar to~\cite{Amhoud2020-ye}. 
Since the skyrmionic beam consists of two orthogonal polarization components, we propagate each component separately but apply the \textit{identical} set of phase screens to both, in a similar manner to~\cite{Wang2025ssfmturb}.
Here, it is assumed that the refractive index fluctuations are polarization-independent and that both components traverse the same optical path.

We utilized a grid of $1024 \times 1024$ pixels with a spacing of $\Delta x = \Delta y = 0.733~\mathrm{mm}$.
This high spatial resolution satisfies the sampling constraints to avoid aliasing~\cite{Schmidt2010-bk}, while simultaneously maintaining sufficiently low discretization errors in the numerical differentiation and integration, required for calculating $\tilde{N}$ at the receiver.
The propagation path is divided into 20 steps of $\Delta z = 50~\mathrm{m}$, satisfying the condition $\Delta z > L_0$, which is required for SSFM. Finally, 20,000 Monte Carlo trials were performed for each configuration to capture the stochastic channel statistics.

\subsection{Intensity-Based Masking}%
\label{subsec:masking_effectiveness}
\begin{figure*}
    \centering
    \subfloat[]{\includegraphics[width=0.28\linewidth]{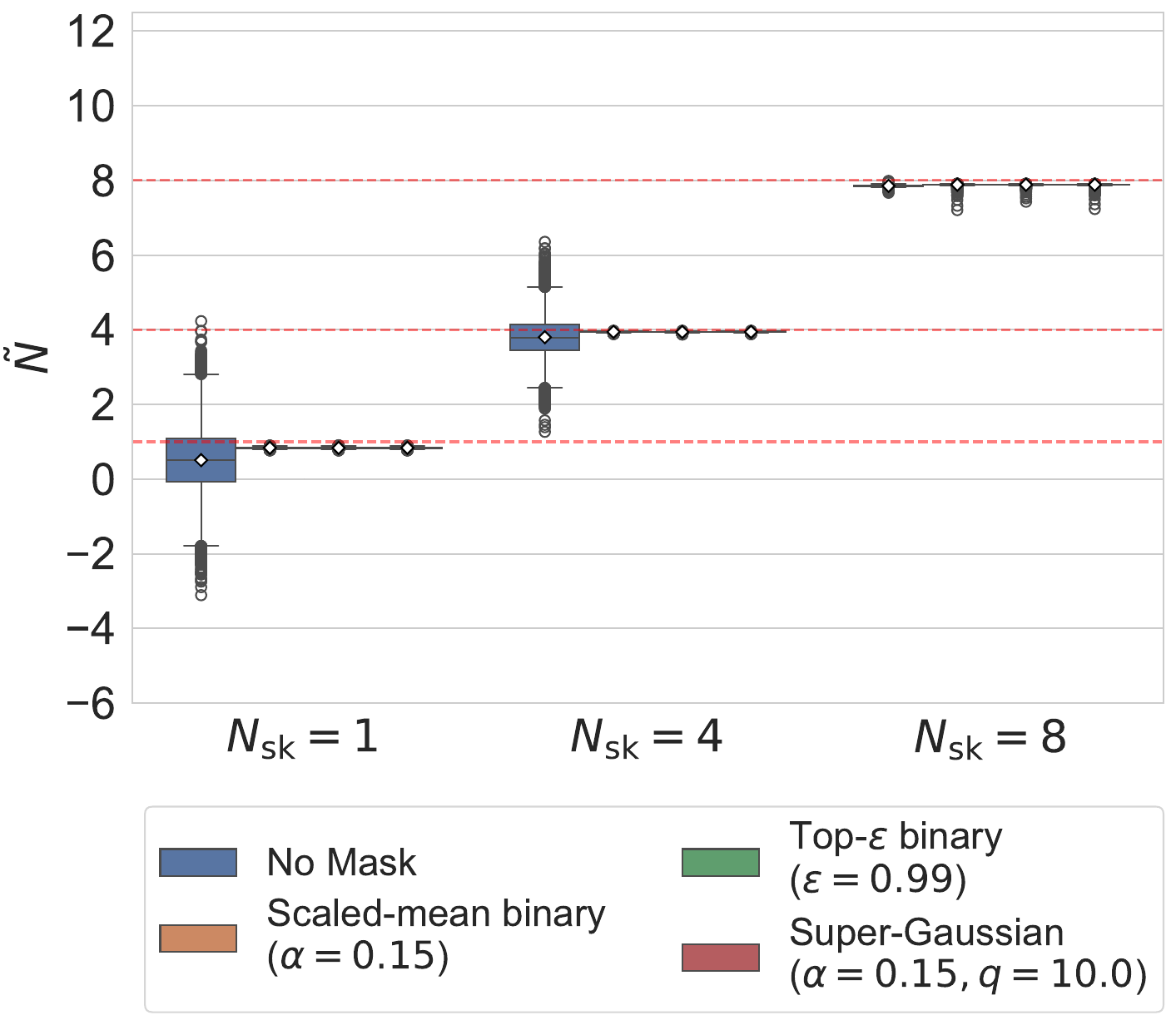}%
        \label{fig:box_plots_a}}
    \hfill
    \subfloat[]{\includegraphics[width=0.28\linewidth]{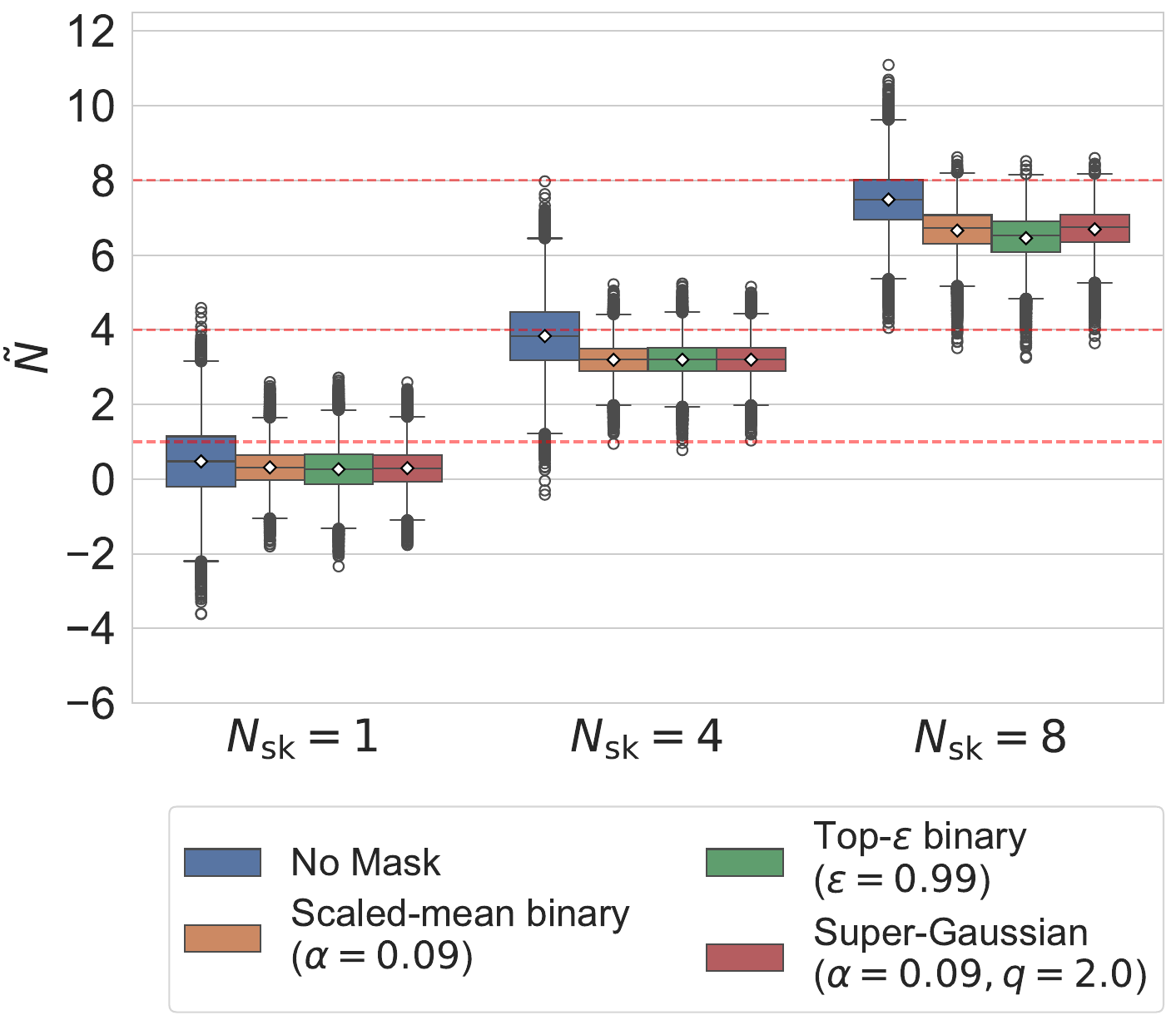}%
        \label{fig:box_plots_b}}
    \hfill
    \subfloat[]{\includegraphics[width=0.28\linewidth]{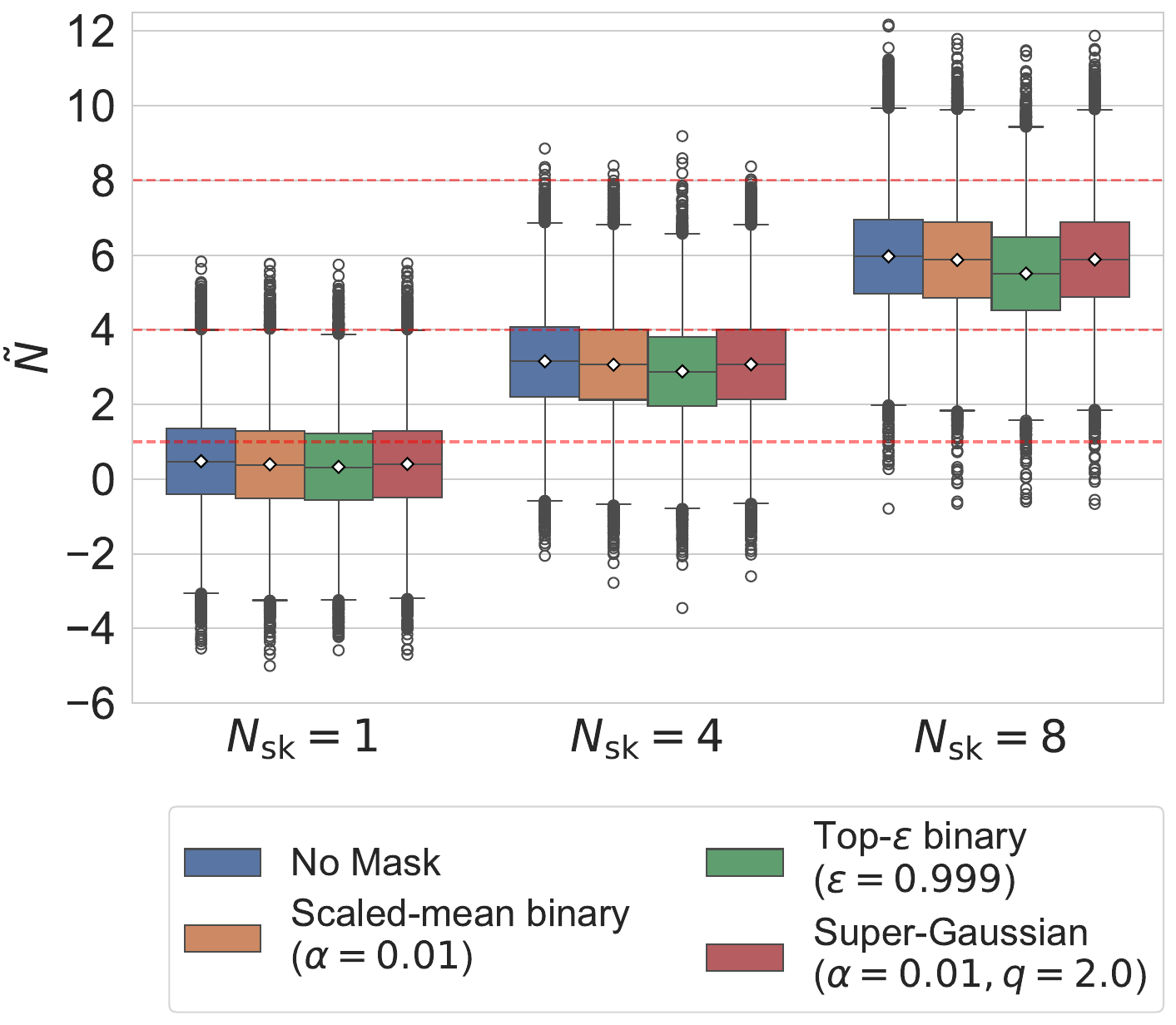}%
        \label{fig:box_plots_c}}
    \caption{Box plots of the estimated skyrmion numbers $\tilde{N}$ for the single positive skyrmion number transmission $N_{\mathrm{sk}} =1$, $4$, and $8$: (a) weak ($C_n^2=1.0\times10^{-15}$), (b) moderate ($C_n^2=2.5\times10^{-14}$), and (c) strong turbulence ($C_n^2 = 1.0 \times 10^{-13}$). Red line represents the transmitted $N_{\mathrm{sk}}$ values.}%
    \label{fig:box_plots}
\end{figure*}
Here, we introduce three specific masking functions, i.e., the \textit{scaled-mean binary}, \textit{top-$\epsilon$ binary}, and \textit{super-Gaussian} ones, to implement intensity-based filtering.
First, let us consider the two binary masks, defined by a global intensity threshold $I_{\mathrm{th}}$, where the mask function is given by
\begin{IEEEeqnarray}{rCL}
    \label{eq:mask_def_br}
    W(\bm{\rho}) =
    \begin{cases}
        1, & \text{if } I(\bm{\rho}) \ge I_{\mathrm{th}}, \\
        0, & \text{otherwise}.
    \end{cases}
\end{IEEEeqnarray}
The \textit{scaled-mean binary} mask sets the threshold proportional to the spatial average intensity $I_{\mathrm{avg}}$, computed specifically over the non-zero intensity regions as follows:
    $I_{\mathrm{th}} = \alpha I_{\mathrm{avg}}$,
where $\alpha > 0$ is a tuning parameter.
Scaling the threshold in this manner ensures robustness against global power fluctuations, since the threshold adapts to the transmission power and channel attenuation.

In the \textit{top-$\epsilon$ binary} mask, $I_{\mathrm{th}}$ is designed, such that the cumulative power reaches a target fraction $\epsilon \in (0, 1]$ of the total received power $I_{\mathrm{total}}$ as follows:
    $\int_{I(\bm{\rho}) \ge I_{\mathrm{th}}} I(\bm{\rho}) \, d\bm{\rho} = \epsilon I_{\mathrm{total}}$.
This approach directly extracts signal-dominant regions by accumulating the highest intensity components.

Finally, the soft \textit{super-Gaussian} mask is formulated by
\begin{IEEEeqnarray}{rCL}
    W(\bm{\rho}) = 1 - \exp\!\left[ -\left( {I(\bm{\rho})}/{\alpha I_{\mathrm{avg}}} \right)^{2q} \right],
\end{IEEEeqnarray}
where $q > 0$ controls the steepness. This function allows a smooth transition for masking, since continuous weights are multiplied to intensities near the threshold.

\Fref{fig:box_plots} shows the skyrmion numbers $\tilde{N}$ estimated with the aid of three masking schemes at the receiver, where a single positive skyrmion number is transmitted, i.e., $N_{\mathrm{sk}}=1$, $4$, and $8$. For reference, the results of a no-mask case are also plotted. We consider the three turbulence levels, i.e., weak ($C_n^2=1.0\times10^{-15}$), moderate ($C_n^2=2.5\times10^{-14}$), and strong ($C_n^2 = 1.0 \times 10^{-13}$).
The masking parameters $\alpha$, $\epsilon$, and $q$ are optimized for each turbulence level $C_n^2$, such that the mean-square error of the masked signal, relative to that of the vacuum propagation case, is minimized.
Observe in \Fref{fig:box_plots} that all of our masking schemes significantly improve the accuracy of $\tilde{N}$ estimation in weak-to-moderate turbulence regimes, compared to the no-mask baseline.
Additionally, for moderate and strong turbulence levels, the downward shifts of the calculated skyrmion numbers $\tilde{N}$ from the transmitted one are observed.
These shifts arise because the masking operation excludes the noise-dominant beam periphery, thus reducing the effective integration area. In the maximum-likelihood detection stage, we account for shifts, thereby minimizing the detrimental effects on the resultant skyrmion-number detection $\hat{N}$.
Since there are no substantial performance differences between the proposed masking schemes in the simulated scenarios, we employ the scaled-mean binary mask for computational simplicity in the rest of this paper.

\section{Performance Analysis}%
\label{sec:Performance_Analysis}
\subsection{Derivation of Theoretical Performance Bounds}
The proposed SkM-based FSO system is modeled based on the discrete memoryless channel (DMC) representation between the transmitted skyrmion number $N_{\mathrm{sk}}$ and the hard-detected one as $\hat{N}$.
Consider the skyrmion number constellation $\mathcal{C} = \{n_1, n_2, \ldots, n_M\}$ with the size of $M$.
The channel is characterized by a transition probability matrix $\mathbf{P} \in \mathbb{R}_{\ge 0}^{M \times M}$, whose $i$th-row and $j$th-column entry $P_{ij}$ represents the probability of detecting symbol $n_j$ given that $n_i$ was transmitted.
$P_{ij}$ is derived from the conditional PDF of $\tilde{N}$ and the decision regions $\mathcal{D}_j$ as
    $P_{ij} = \Pr(\hat{N} = n_j \mid N_\mathrm{sk} = n_i)
    = \int_{\mathcal{D}_j} f(x \mid n_i) \, dx$.

Based on $\mathbf{P}$, the theoretical performance metrics, including the channel capacity, SER, and BER, are derived.
Let $p(n_i)$ denote the \textit{a priori} probability of the transmitted skyrmion number $n_i$. The channel capacity $C$ quantifies the maximum achievable information rate of the system and is formulated by maximizing the mutual information $I(N_\mathrm{sk}; \hat{N})$ over all possible input distributions $p(n_i)$ as follows:
    $C = \max_{ p(n_i) } I(N_\mathrm{sk}; \hat{N})$,
where the maximization is carried out numerically using the Arimoto-Blahut algorithm, yielding the capacity-achieving input distribution $p^*(n_i)$.

The theoretical SER $P_s$ is formulated by 
    $P_s = \sum_{i=1}^{M} \sum_{\substack{j=1 \\ j \ne i}}^{M} P_{ij} \, p(n_i)$.
Similarly, the theoretical BER $P_b$ is given by 
    $P_b = \frac{1}{\log_2 M} \sum_{i=1}^{M} \sum_{\substack{j=1 \\ j \ne i}}^{M} d_H(\bm{c}_i, \bm{c}_j) P_{ij} \, p(n_i)$,
where $d_H(\bm{c}_i, \bm{c}_j)$ denotes the Hamming distance between the bit sequences $\bm{c}_i$ and $\bm{c}_j$, which correspond to the modulated skyrmion numbers $n_i$ and $n_j$, respectively.
We employ Gray coding for bit mapping to minimize bit errors between adjacent symbols.
In the subsequent analysis, we use the capacity-achieving distribution $p^*(n_i)$ for evaluating $P_s$ and $P_b$.

\begin{table}
    \caption{BER, SER, and Channel Capacity for Optimal Constellations under Three Turbulence Conditions}%
    \label{tab:performance_summary}
    \centering

    \begin{threeparttable}
        \scriptsize
        \begin{tabular}{@{}llccc@{}}
            \toprule
            \textbf{Turbulence Strength} $C_n^2$ & \textbf{M} & \textbf{SER}          & \textbf{BER}          & \textbf{Capacity} \\
            \midrule

            \multirow{4}{*}{Weak ($1.0\times10^{-15}$)}
                                                 & 4          & $6.77\times10^{-117}$ & $3.39\times10^{-117}$ & 2.00                     \\
                                                 & 8          & $8.03\times10^{-66}$  & $2.68\times10^{-66}$  & 3.00                     \\
            \addlinespace
                                                 & 16         & $4.57\times10^{-18}$ & $1.14\times10^{-18}$ & 4.00              \\

            \midrule

            \multirow{4}{*}{Moderate ($2.5\times10^{-14}$)}
                                                 & 4          & $7.79\times10^{-4}$   & $3.89\times10^{-4}$   & 1.99                     \\
            \addlinespace
                                                 & 8          & $6.44\times10^{-2}$   & $2.16\times10^{-2}$   & 2.61                     \\
            \addlinespace
                                                 & 16        & $3.25\times10^{-1}$  & $8.56\times10^{-2}$  & 2.73              \\
            \midrule

            \multirow{4}{*}{Strong ($1.0\times10^{-13}$)}
                                                 & 4          & $1.12\times10^{-1}$   & $5.63\times10^{-2}$   & 1.42                     \\
            \addlinespace
                                                 & 8          & $3.07\times10^{-1}$   & $1.20\times10^{-1}$   & 1.52                     \\
            \addlinespace
                                                 & 16        & $5.54\times10^{-1}$  & $2.01\times10^{-1}$  & 1.56              \\
            \bottomrule
        \end{tabular}
    \end{threeparttable}
\end{table}

\subsection{Constellation Optimization}
Given a constellation size $M$, the constellation is optimized by the exhaustive search over all possible subsets of the available skyrmion number set $\mathcal{S}$ with cardinality $M$. We adopt the channel capacity as the objective function to be maximized.
Hence, the optimal constellation $\mathcal{C}^*$ may be written as
    $\mathcal{C}^* = \argmax_{\mathcal{C} \subset \mathcal{S}, |\mathcal{C}| = M} C(\mathcal{C})$,
where $C(\mathcal{C})$ denotes the channel capacity computed for the constellation subset $\mathcal{C}$.

\subsection{Results and Discussion}
In \Tref{tab:performance_summary}, we present performance results for the proposed scheme in terms of the derived SER, BER, and channel capacity under weak, moderate, and strong turbulence conditions, along with the optimized skyrmion number constellation. The constellation size is given by $M=4$, $8$, and $16$. 
Under the weak turbulence scenario, the proposed SkM scheme exhibits good performance, with the capacity approaching the transmission rate $\log_2M$ for each $M$. 
This confirms the robustness of the proposed SkM scheme under the weak turbulence.
Moreover, in the moderate turbulence scenario, there are non-zero BER and SER, where the performance degradation increases upon increasing the constellation size $M$. 
Note that, with a powerful channel coding scheme, error-free performance may be attainable when the transmission rate is lower than the channel capacity. Additionally, increasing the constellation size from $M = 8$ to $16$ yields only a marginal capacity gain (from 2.61 to 2.73 BPCU), due to increased overlap between the distributions.
Under the strong turbulence scenario, channel capacity decreases further. 
The BER exceeds $10^{-2}$ even for $M=4$, and the channel capacity saturates at approximately 1.5 BPCU regardless of $M$.

\section{Conclusion}%
\label{sec:Conclusion}
In this paper, we proposed a novel FSO communication system that utilizes optical skyrmions, where the skyrmion number is exploited as a new degree of freedom for IM.
We derived theoretical performance bounds for the proposed SkM scheme, including the channel capacity, SER, and BER.
Finally, our performance analysis based on the DMC framework confirmed that the proposed system exhibits near-ideal robustness under weak turbulence and supports high-order modulation in moderate regimes.

\section*{Acknowledgement}
This work was supported in part by the JST FOREST (Grant JPMJFR2127), in part by the JST ASPIRE (Grant JPMJAP2345), in part by the JST CREST (JPMJCR19T1), and in part by the JSPS KAKENHI (Grant 24K21615).

\bibliographystyle{IEEEtran}
\bibliography{IEEEabrv,skyrmion_paper}
\end{document}